# U solubility in Earth's core

Xuezhao Bao[*] and Richard A. Secco

Department of Earth Sciences, University of Western Ontario, London, Ontario, Canada N6A 5B7

**Abstract**

Uranium is the most important heat producing element in the Earth. The presence of an appreciable amount of U in Earth's core would have an important influence on geodynamics. In this study, the solubility of U in Fe-10wt% S and in Fe-35wt% S was measured by partitioning experiments with a mixture of peridotite, uraninite, Fe and FeS powder at pressure (P) of 0-9 GPa and temperature (T) of 1500-2200 $^{o}$C. Comparisons with the run products containing pure Fe as the metal phase in our previous study and re-analysis of run products were made in this study. We found that in all run products, including Fe-10wt% S, Fe-35wt% S and pure Fe groups, the solubility and partitioning of U in the pure metal or metal-sulfide phase relative to the silicate phase ($D_U$) increases with increasing P and T. With a molten silicate phase, $D_U$ is generally 3-6 times larger than with a solid silicate phase. While $D_U$ has a positive dependence on S concentration of the metal-sulfide phase, there is a negative correlation between Ca and U. According to our calculations based on these experimental results, if the core has formed from a magma ocean at a P of 26 GPa at its base and the core contained 10wt% S, then it could have incorporated at least 10 ppb U. Alternatively, if the core formed by percolation and contained 10wt% S, then it could have incorporated 5-22 ppb U. The geophysical implications of U in Earth's core are discussed.

---
[*] Corresponding author. 1-519-661-4079; Fax: 1-519-661-3198
E-mail: xuezhaobao@hotmail.com (Xuezhao Bao); secco@uwo.ca (Richard A. Secco)

1. INTRODUCTION

Models of energy sources for the driving force of Earth's heat engine involve ongoing growth of the inner core by cooling, releasing gravitational potential energy and latent heat of crystallization (Stacey, 1992). These energy sources, however, are generally capable of changing only gradually in one direction over time. The Earth's energy output, by contrast, appears to vary in intensity frequently, but quite irregularly. For instance, in the recent 350 Ma of Earth's history, there were two geomagnetic superchrons, namely the 320-250 Ma normal polarity and 124-83 Ma reversed polarity. During these stages, the geomagnetic field stopped its reversal activity, and may have had the highest intensity (Tarduno et al., 2001), but this coincided with many large volcanic eruptions, mantle plume activities and biota extinction events (Courtillot and Besse, 1987; Prevot et al., 1990; Larson and Olson, 1991). This implies a close relationship between large-scale geological activities and the energy production in and output from the core.

Herndon (1996) put forward a hypothesis that considers the energy from nuclear fission of U in the Earth's inner core to be the main energy source to maintain the geomagnetic field. Labrosse et al. (2001) modelled the timing and rate of inner core solidification and Anderson (2002) considered the energy balance at the Core Mantle Boundary (CMB) and both concluded that there is a need for some amount of radiogenic heating in the core. Olson (2006) recently suggested the need for additional core heat source(s), including radioactive heating, to explain the high temperature predicted prior to inner core formation. $^{40}$K is a possible radioactive source in planetary cores based on high pressure(P)-temperature(T) experimentation of K solubility in the Fe sulfide (Murthy et al., 2003). Recent observation of geoneutrinos released from U and Th decay chains suggested that the radiogenic heat power that comes only from U and Th within the Earth can reach an upper limit of 60 TW with a central value at 16 TW (Araki et al., 2005). Geochemical modeling predicts that the radiogenic heat from U and Th within the Earth's crust and mantle is only 16 TW (Verhoogen, 1980). Therefore, the wide range of upper limit values suggested for U and Th heating from geoneutrino observation admits the potential for U and Th inclusion in Earth's core.

The seismologically inferred densities of Earth's outer and inner cores are lower than the density of pure Fe at core P, T conditions. Therefore, it is often suggested that approximately 10% and 3% of light elements must be present in the outer and inner cores, respectively, to account for these density discrepancies (Birch, 1952; Stixrude and Wasserman, 1997). Sulphur, silicon and oxygen are thought to be the most possible candidates (e.g. Poirier, 1994; McDonough, 2003). In a previous experimental study (Bao et al., 2006), a positive correlation between the concentration of Si and the solubility of U in the Fe phase was found with a mixture of uraninite, peridotite and metallic iron powder. In the set of experiments described in this study, the influence of S on the solubility of U in the metal phase is investigated by using Fe-10wt% S and Fe-35wt% S as the metal phase components. These results are compared with those achieved from previous run products with a metal phase of pure Fe.

## 2. EXPERIMENTAL DETAILS

### 2.1 The starting material

The starting material in each experiment was produced by mixing and grinding Fe and FeS (Aldrich Chem. Co., 99.99% pure), natural peridotite (Bay of Islands ophiolite suite, Newfoundland, Canada) and uraninite (containing approximately 80 wt% $UO_2$, Goldfields, Saskatchewan, Canada). The mixture was composed of ~40 wt% metal phase of Fe-10wt% S or Fe-35wt% S, approximately 54 wt % silicate and approximately 6 wt% $UO_2$ (uraninite). As in the pure Fe study (Bao et al., 2006), sufficient uraninite was added to the starting mixture in order to create conditions whereby the concentration of U in the metal phase might exceed the detection limit of the analysis instruments.

### 2. 2 high pressure and temperature experiments

The high-pressure experiments were carried out in a Walker-type module high-pressure device (Walker et al., 1990). The pressure calibration of the press was described in previous work (Secco et al., 2001). The pressure cell was a cast, pre-gasketed, MgO octahedron with edge length of 18, 16 and 14 mm for tungsten carbide cubes with truncations of 8, 6, or 4 mm, respectively. Approximately 40 to 50 mg of sample powder was loaded into a boron nitride (BN) or graphite capsule, which occupied the center of the pressure cell. The capsule was surrounded by a cylindrical Nb or graphite furnace, which was thermally insulated from the MgO octahedron by a zirconia sleeve. An MgO plug and sleeve were used to separate the sample capsule and Fe conduction rings at the ends. The temperature was measured by a W-3%Re/W-25%Re thermocouple situated directly above the capsule and pressure correction for the emf was not applied. All assembly components were fired at 200 $^o$C for one hour before loading. For the lowest pressure 3GPa runs, pressure was slowly increased to the target pressure at a pressurization rate of approximately 2GPa/hr. For all other runs, pressure was increased rapidly to 2.5GPa followed by a pressurization rate of approximately 2GPa/hr to the target pressure. After being compressed to the desired pressure, the temperature was then raised at 100 $^o$C/min to the desired run temperature and then held there for 3 to 73 min. Temperature gradients were measured in prior experiments, and values of 100$^o$C/mm were obtained at the maximum temperature in this study.

After completion of the experiment, the entire sample assembly was mounted in epoxy, sectioned through the center of the sample and then polished for Laser Ablation Inductively Coupled Plasma Mass Spectrometry (LA-ICP-MS) and Electron Microprobe (EMP).

2.3 Electron Microprobe

The major elements and light elements (C, N and O) in the silicate, metal and uraninite phases were analyzed by electron microprobes. For major element analyses, electron microprobes at the University of Western Ontario (Western), University of Manitoba (Manitoba), University of Toronto (Toronto), and Canada Centre for Mineral and Energy Technology (CANMET) at Ottawa (Ottawa) were used. The electron microprobe at Toronto was mainly used to analyze U in metal phases and the microprobe at CANMET was used for comparison since both laboratories have U standard materials. Similarly, uraninites were analyzed at Manitoba. For the light element analyses, the electron microprobe from McGill University was used.

Most analyses were conducted using a Jeol 8600 EMP at Western. An electron beam with a voltage of 15 kV and a current of 15 nA was used. Raw data were reduced using the ZAF correction built into Tracor Northern automation system. The beam size is 1 μm in this study. The depth of the electron beam penetration is 1 to 1.5 μm. Several samples were reanalyzed using other EMPs mentioned above at similar conditions. Usually for major elements, the accuracy of the microprobe is approximately ± 1.5 %. The detection limits under the conditions used in this study were: Mg ~300 ppm; Si ~210 ppm; Ca 260 ppm; C, N and O: > 700 ppm.

2.4 Laser Ablation Inductively Coupled Plasma Mass Spectrometry

Analysis of U and other trace elements were conducted mainly at the LA–ICP–MS facility in McGill University using procedures and conditions described in Wheeler et al. (2006). With an optical system in a microscope, a laser beam is focused on the surface of a sample. The laser beam size can be adjusted from 10 to 40 μm in diameter according to exposure area. Before LA-ICP-MS sampling, pre-ablation was done to get a clean surface. A glass from the National Institute of Standards and Technology (NIST), NIST 610 were used as the external standard. Usually, the standard was run every four to six experiments. Fe and Si from the electron microprobe analysis were used as the internal standards. Every phase in each experiment was measured two to ten times depending on its exposure area. The measured data were reduced using the GLITTER LA-ICP-MS processing software, written by GEMOC, MacQuarrie University, Australia.

Several samples were analyzed at the LA–ICP–MS facility in the Great Lakes Institute for Environmental Research at the University of Windsor using procedures and conditions described in Gagnon et al. (2003) and our previous study (Bao et al., 2006). The detection limits are: U 0.01 ppm, Th 0.04 ppm; Mg 4.76 ppm and Ca 102 ppm.

3. RESULTS AND DISCUSSION

3.1 P, T range

Fig.1 shows the P and T range of our experiments. The peridotite consists mainly (60-80 wt %) of Mg-olivine and therefore the forsterite melting line was added to the Fig.1 as a reference. Fe-10wt% S and Fe-35wt% S melting lines were also added to the Fig.1 since they are the metal phase in the starting material by mixing Fe and FeS. Most of the experimental temperatures were between the Mg-olivine and metal phase melting lines, but are closer to the melting line of Mg-olivine. Therefore, the metal sulfides and metal phase were liquid during the experiments. The large (> 100 μm) and clear metal crystals in these run products as shown in Fig.2 support this. Compared with the starting powder (< 50 μm), the silicate phase also has clear crystals larger than 100 μm in size in most run products (Fig.2), which show that the run temperatures reached the re-crystallization temperature. The temperatures of runs 260 and 262 were estimated from their power values. In addition, the very large (>600 μm) metal phases in runs 262 and 260 indicate that their metal phases were melted during experiments, therefore their run temperatures should be higher than the Fe melting temperatures (namely > 1910 °C and > 1953 °C, respectively).

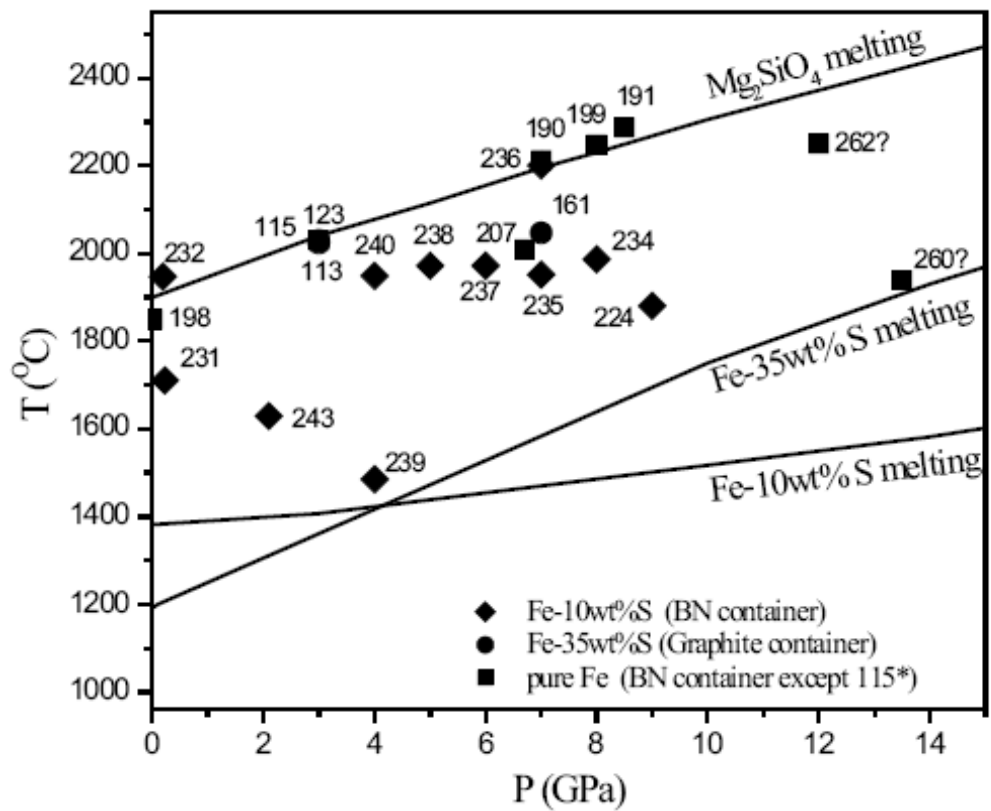

**Fig.1 The T and P conditions of the experiments. The T values are the highest experimental T in most runs and as measured by thermocouples. The T of run 236 was determined by the relation between power and temperature. The T of runs 260 and 262 are determined from their power value and phase transformation of their metal phases, therefore, their T values are uncertain and marked with "?" signs. The reference olivine melting line is based on the data of Davis & England (1964), and Ohtani & Kumazawa (1981); and metal-sulfide melting lines is based on the data of Fei et al. (1997), and Brett & Bell (1969). * Graphite container was used in run 115.**

3.2 Composition analysis, results and discussion

The major compositions of the metal phases analyzed by electron microprobe are listed in Table 1. The LA-ICP-MS line scans for U content in Fe-10wt% S, Fe-35wt% S and Fe are shown in Figs. 3-5. The variation of counts in these spectra in each figure was caused by back and forth sampling in the same track but at different depth. Fig.3 shows the effect of pressure on the solubility of U in Fe-10wt% S between 1947 and 2200 °C. The amount of U entering the metal-sulfide phase is indistinguishable from the background up to a pressure of 4 GPa. With increasing pressure above 4 GPa, U solubility increases significantly up to a P of 7 GPa as indicated by ordinate axis scale. A similar effect of pressure on U content in Fe-35wt% S and Fe is shown in Figs.4 and 5. In fig.4, the two upper plots are at very similar run condition and the data indicate the reproducibility of the results. As shown in Fig.1, the experimental temperatures of runs 113, 123, 232 and 236 in the Fe-10wt% S group and runs 115, 190, 191 and 199 in the pure Fe group are close to or above the silicate phase (Mg-olivine) melting line, but only the high P runs 190, 191, 199 and 236 have high U content. This is consistent with the pure Fe run products in the previous experiments on pure Fe (Bao et al, 2006), namely when P ≥ 7GPa and T > silicate melting T, U concentration in the metal phase is several times higher than that of the low T run products. Run 161 in the Fe-35wt% S group also has very high U content, which may reflect that S can accelerate U solubility in metal phase. This is discussed in detail later.

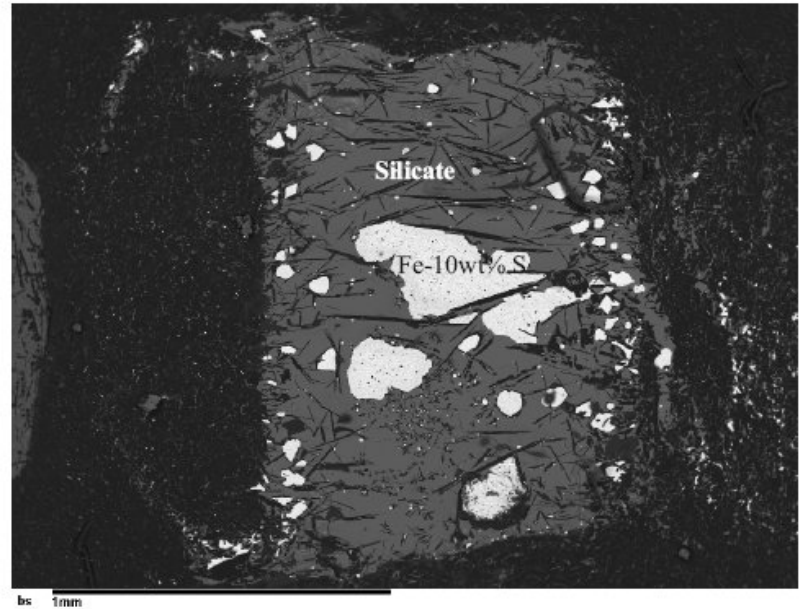

**Fig.2  Recovered sample from run 238 showing the texture of the Fe-10wt% S (white areas) and silicate phases (grey areas).**

**Table 1 Average composition in the metal phases analyzed by Electron Microprobe (wt %)**

| Run # | Fe | S | Si | Mg | Total* |
|-------|-------|-------|------|------|--------|
| 113 | 62.28 | 25.72 | N/A | N/A | 88.12 |
| 115 | 96.38 | 0.05 | 0.01 | 0.00 | 96.44 |
| 123 | 62.71 | 31.91 | 0.01 | N/A | 94.64 |
| 161 | 60.80 | 35.11 | 0.01 | 0.01 | 95.93 |
| 190 | 94.59 | 0.01 | 0.31 | 0.00 | 94.91 |
| 191 | 94.12 | 0.03 | 0.51 | 0.00 | 94.66 |
| 199 | 93.39 | 0.02 | 0.40 | 0.00 | 93.82 |
| 232 | 89.41 | 0.24 | 1.45 | 0.01 | 91.10 |
| 234 | 82.65 | 12.20 | 2.14 | N/A | 97.49 |
| 235** | 82.85 | 11.37 | 0.07 | N/A | 97.08 |
| 236 | 91.17 | 4.52 | 0.14 | N/A | 96.22 |
| 237 | 87.85 | 7.95 | 0.11 | N/A | 95.50 |
| 238 | 81.77 | 10.78 | 0.26 | N/A | 93.24 |
| 239 | 81.41 | 10.71 | 0.01 | N/A | 92.51 |
| 240 | 91.62 | 2.09 | 0.15 | N/A | 94.32 |
| 243 | 87.07 | 2.68 | 2.87 | N/A | 92.97 |
| 260 | 83.71 | 0.02 | 3.18 | 0.01 | 86.92 |
| 262 | 97.12 | 0.06 | 1.29 | 0.00 | 98.46 |

**\* The metal phases may contain other elements, such as light elements B, N and O coming from container and starting materials.**
**\*\* Light elements were analyzed in Run 235 (wt %): O 0.70; B 0.05; N 1.55.**
**We didn't obtain ideal electron microprobe data from runs 198, 231 and 224 because of unfavorable sample conditions for EMP analysis. Therefore, average Fe and Si concentrations calculated from the starting material group they belonged to, was used for LA-ICP-MS analyses in these three runs.**

The LA-ICP-MS results from McGill for the quantitative analysis of U were used for this set of experiments. However, in order to check machine system error(s), several pure Fe run products from the pure Fe experiment analyzed using the Windsor LA-ICP-MS were re-analyzed at McGill and the comparison is given in Tables 2. The metal phase in runs 115, 190, 191 and 199 were pure Fe and were analyzed with the Windsor LA-ICP-MS in our previous study. Comparing U concentration values acquired on the McGill LA-ICP-MS, it was found that the U concentration in the lower P run 115 is less than 15 ppm in both instruments: 0.6 ppm (Windsor) versus 14.62 ppm (McGill). This indicates that the difference between the two machines is 14 ppm in this run product (Table 2). Similarly, the other low P ($\leq$ 4 GPa) run products, including runs 198, 231, 232, 243, 239, 240, 113 and 123, in Tables 2 and 3 and also plotted in Fig.6, also have very low and similar U concentration (most of them are < 20 ppm, Table 2). This indicates that there is a difference between the two instruments, but it is in the range of 20 ppm. In the higher P run products, the U concentrations in run 190 are almost the same from the two instruments: 418.5 ppm (Windsor) versus 414.7 ppm (McGill) with a difference between them of only 1%. For run 191, U concentration is 796 ppm from Windsor, and 997.9 ppm from McGill with an error between them of 20%. Run 199 has the largest difference between the two instruments: U from Windsor is 624.6 ppm, but 2306.6 ppm from McGill. This difference may originate from the uneven distribution of U in Fe phase of run 199. The LA-ICP-MS spectra of run 199 in Fig.5 also show that uneven distribution of U in Fe phase is apparent in some samples. In conclusion, the two instruments may have some system errors, but they are not large enough to influence the conclusions.

The solubility of Mg in Fe phases is very small and is generally < 0.06 wt % or 0.1 wt% in MgO. Therefore, Mg < 0.06 wt% in metal phase is used as an upper limit to choose useful LA-ICP-MS data as discussed in detail in our previous study (Bao et al., 2006, Table 3).

The concentration of U in the metal-sulfide and the pure Fe is shown in Fig.6. In general, an increase in U solubility is observed with increasing pressure. For the Fe-10wt% S group, the experimental temperatures serve as a means for the division into two sub-groups, those with T above the silicate melting line and those with T below the silicate melting line. In the first sub-group, the temperature of runs 232 and 236 are slightly higher than the Mg-olivine melting (Fig.1), which means both silicate and metal phases were melted during the experiments, but only the higher pressure run 236 at 7 GPa had a high concentration of U. In fact, it had the highest U concentration among all the Fe-10wt% S run products, which is consistent with the results on pure Fe, namely when P ≥ 7 GPa and run T > silicate melting line, U concentration is 4-6 times larger that the lower T run products. In the second sub-group, where the experimental T was below silicate melting, the results of nine runs clearly indicate that U concentration increases stably with increasing P, which is also consistent with the pure Fe run products reported earlier (Bao et al., 2006). It is convenient.

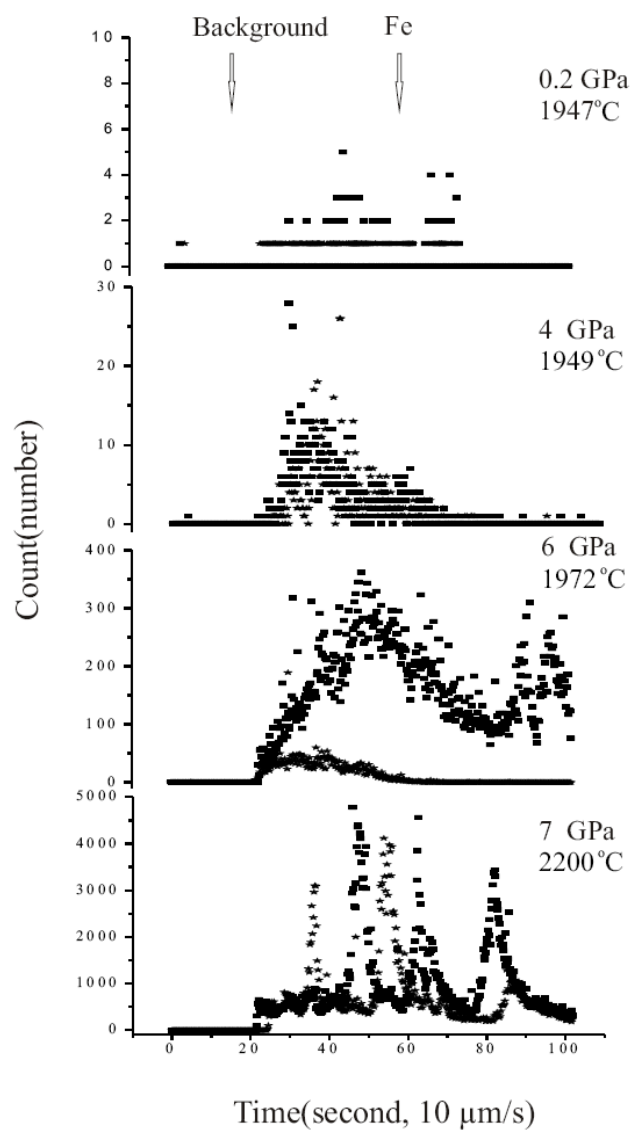

**Fig.3  Spectra of U in the Fe-10wt% S phases at different pressures analyzed by LA-ICP-MS. Run numbers from top to bottom: 232, 240, 237 and 236. Different symbols in the spectra indicate different analysis lines, which also applies to Figs. 4 and 5.**

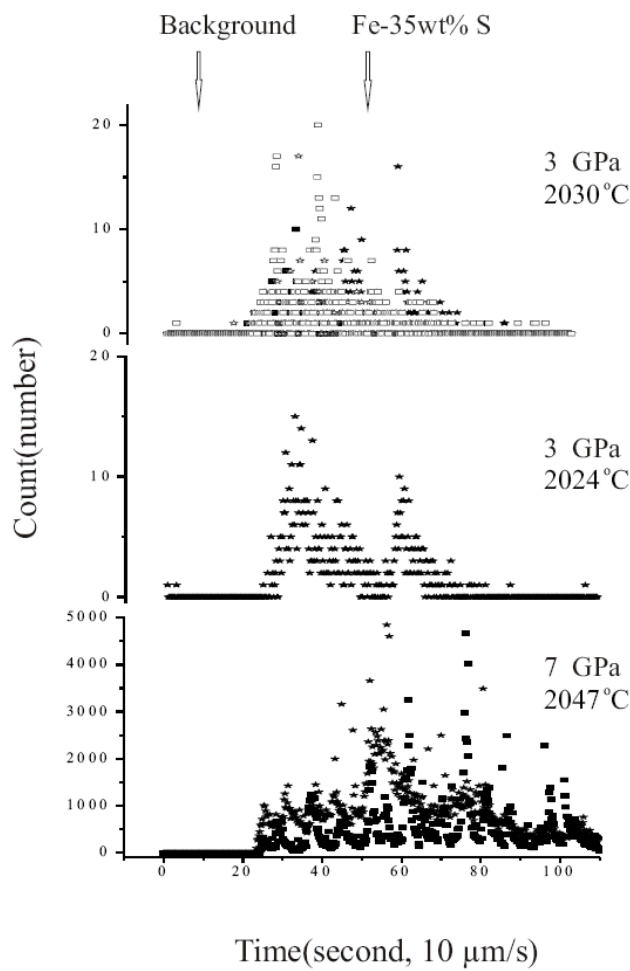

**Fig.4  Spectra of U in the Fe-35wt% S phases at different pressures analyzed by LA-ICP-MS. Run numbers from top to bottom: 113, 123 and 161.**

In order to compare any effect of sulphur on the solubility and partitioning of U in the metal phase, 4 new Fe products (runs 198, 207, 260 and 262) were analyzed and another 4 run products (runs 115, 190, 191, 199) were re-analyzed with the McGill LA-ICP-MS to allow direct comparison with the sulfide run products. The pure Fe runs were also divided into the same two sub-groups on the basis of temperature as described for the Fe-10wt% S runs. The runs at temperatures above silicate melting are runs 115, 190, 191 and 199. Except for the low P run 115 with 7 ppm U, the other three with P ≥ 7 GPa had a U concentration higher than 350 ppm. On the contrary, the U concentrations in each of the run products with temperatures below silicate melting, runs 198 and 207 are lower than the former sub-group. But they (including run 260) still trend in the same way, namely U concentration increases with increasing pressure. The pattern of the Fe group further supports the analysis results and conclusion arrived at on the basis of the experimental run products analyzed using the Windsor LA-ICP-MS. A very high U concentration of 8686 ppm for run 262 as shown in Fig.6 may originate from the high experimental P (12GPa) and T, but it can not be assigned to any sub-group because the accuracy of T is not known because a relation between T and furnace power in the 4 mm pressure cell has not been established.

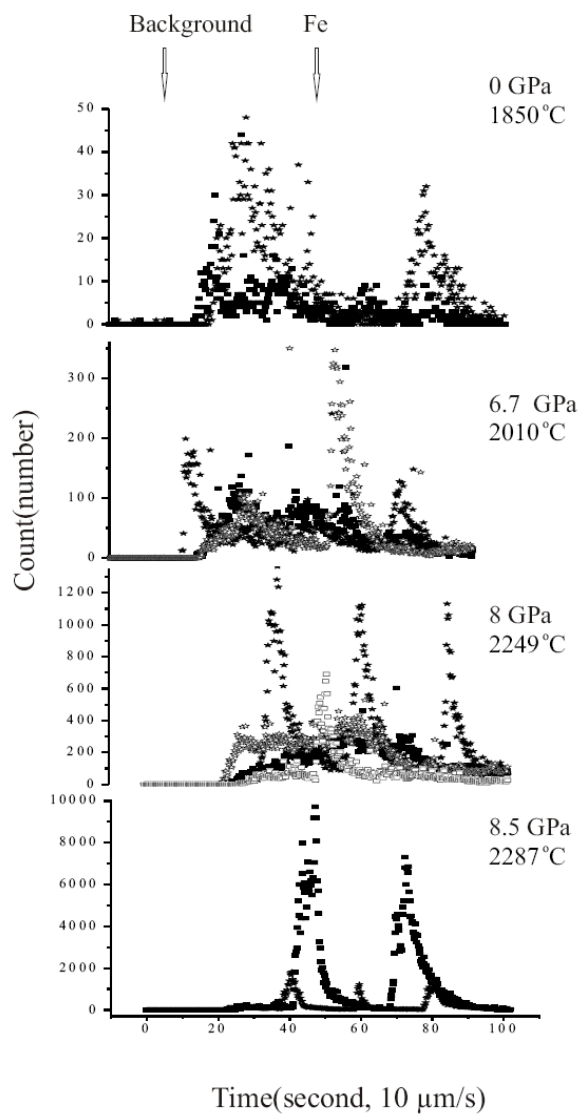

**Fig.5 Spectra of U in the Fe phases at different pressures analyzed by LA-ICP-MS. Run numbers from top to bottom: 198, 207, 199 and 191.**

**Table 2 U concentration in the metal phase with a low experimental P and the comparison of U concentrations between LA-ICP-MS analyses at McGill and Windsor**

| Run # | Metal phase | P (GPa) | U from ICP (ppm) | |
|-------|-------------|---------|------------------|---|
| | | | McGill | Windsor |
| 198 | Fe | 0 | 2.21 | |
| 232 | Fe-10wt% S | 0.2 | 1.98 | |
| 231 | Fe-10wt% S | 0.2 | 16.5 | |
| 243 | Fe-10wt% S | 2 | 38.88 | |
| 123 | Fe-35wt% S | 3 | 13.2 | |
| 113 | Fe-35wt% S | 3 | 6.41 | |
| 200 | Fe | 3 | | 0.7 |
| 115 | Fe | 3 | 14.62 | 0.6 |
| 204 | Fe | 3.5 | | 53.4 |
| 240 | Fe-10wt% S | 4 | 13.69 | |
| 239 | Fe-10wt% S | 4 | 30.32 | |
| 190 | Fe | 7 | 414.86 | 418.5 |
| 199 | Fe | 8 | 2306.59 | 624.6 |
| 191 | Fe | 8.5 | 997.91 | 796 |

The temperatures of runs 113, 123 and 161, which belong to the Fe-35wt% S group, are less than or close to silicate melting. The pressure of runs 113 and 123 was 3 GPa and their metal phases only contained 6-13ppm U. However, the metal phase of run 161 at 7 GPa contains 7156 ppm U, the highest U concentration in all of the metal sulfide phases, including Fe-10wt% S and Fe-35wt% S groups, which also supports a positive pressure dependence of U solubility in metal-sulfides discussed above.

From the data of pure Fe, Fe-10wt% S and Fe-35% S in Fig.6, the following general trends are identified: a) run temperatures above silicate melting result in higher U concentration in the metal or metal-sulfide phase than run temperatures below silicate melting; b) pure Fe data at temperatures above silicate melting overlap the Fe-10wt% S data at temperatures below silicate melting (run 199 is a notable exception); c) the presence of S appears to enhance the solubility of U in the metal phase at high pressures and approximately equivalent temperatures.

**Table 3 LA-ICP-MS analysis result and $D_U$ values**

| P(GPa) | No | Mg (wt%) | U in Fe,ppm | U in silicate,ppm | $D_U$ | Ca(wt%) | Si(wt%) |
|---|---|---|---|---|---|---|---|
| 0 | 198 | 0.007 | 2.21±0.43 | 30858.00±483.70 | 0.00007±0.00001 | 1.06±0.08 | |
| 3 | 115 | 0.036 | 14.62±1.99 | 23622.68±770.55 | 0.00062±0.00009 | 0.57±0.03 | |
| 6.7 | 207 | 0.015 | 271.13±7.80 | 57452.32±7234.00 | 0.00472±0.00061 | 0.56±0.02 | |
| 7 | 190 | 0.021 | 414.86±23.50 | 20980.19±1634.98 | 0.01977±0.00191 | 0.67±0.03 | |
| 8 | 199 | 0.030 | 2306.59±62.20 | 25842.37±4792.70 | 0.08926±0.01673 | 1.84±0.05 | |
| 8.5 | 191 | 0.011 | 997.91±50.99 | 26392.43±2373.87 | 0.03781±0.00391 | 0.70±0.04 | |
| 12 | 262 | 0.076 | 8686.80±78.00 | 118189.05±724.00 | 0.07350±0.00087 | 1.26±0.05 | |
| 13.5 | 260 | 0.061 | 1034.39±14.69 | 68429.91±541.00 | 0.01512±0.00025 | 2.25±0.09 | |
| | | | **U in Fe10wt% S** | | | | |
| 0.2 | 232 | 0.004 | 1.98±0.18 | 18726.55±5128.53 | 0.00011±0.00003 | 0.40±0.03 | 0.57±0.01 |
| 0.2 | 231 | 0.022 | 16.5±1.78 | 17324.26±169.79 | 0.00095±0.00010 | 1.99±0.15 | 2.61±0.04 |
| 2 | 243 | 0.019 | 38.88±2.39 | 10052.69±27.60 | 0.00387±0.00024 | 1.57±0.07 | 4.00±0.05 |
| 4 | 240 | 0.015 | 13.69±0.54 | 16765.56±111.86 | 0.00082±0.00003 | 0.43±0.02 | 0.56±0.01 |
| 4 | 239 | 0.006 | 30.32±6.40 | 17330.48±42.34 | 0.00175±0.00037 | 0.26±0.01 | 0.37±0.00 |
| 5 | 238 | 0.053 | 406.16±15.68 | 21865.52±454.70 | 0.01858±0.00081 | 1.18±0.09 | 1.45±0.03 |
| 6 | 237 | 0.017 | 436.47±17.63 | 29416.80±850.20 | 0.01484±0.00074 | 0.70±0.03 | 0.99±0.02 |
| 7 | 235 | 0.005 | 573.13±18.96 | 14399.49±168.67 | 0.03980±0.00140 | 0.40±0.02 | 1.06±0.05 |
| 7 | 236 | 0.011 | 1586.86±23.76 | 11830.41±270.99 | 0.13413±0.00367 | 0.35±0.02 | 0.89±0.03 |
| 8 | 234 | 0.016 | 945.88±32.90 | 18726.55±515.18 | 0.05051±0.00224 | 0.52±0.03 | 1.37±0.02 |
| 9 | 224 | 0.052 | 1074.91±67.37 | 29554.02±189.00 | 0.03637±0.00229 | 2.02±0.21 | 6.88±0.21 |
| | | | **U in Fe35wt% S** | | | | |
| 3 | 123 | 0.003 | 13.2±0.24 | 20317.38±72.43 | 0.00065±0.00001 | 0.22±0.01 | 0.31±0.00 |
| 3 | 113 | 0.005 | 6.41±0.37 | 13371.53±36.86 | 0.00048±0.00003 | 0.41±0.02 | 0.50±0.01 |
| 7 | 161 | 0.038 | 7155.87±1428.34 | 97386.63±1169.37 | 0.07348±0.01469 | 0.41±0.01 | 0.52±0.01 |

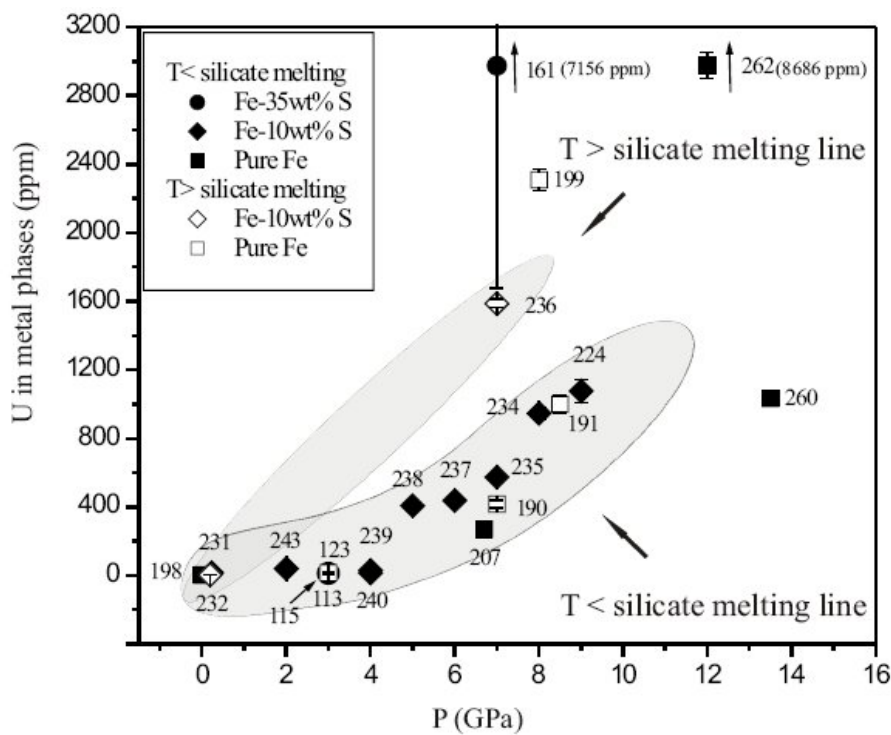

**Fig.6  U concentration in Fe and metal-sulfide phases as a function of pressure. The containers for Fe-35wt% S group are graphite and the others are BN. The lower shading region outlines the Fe-10wt% S group with a T below silicate melting. Errors are 2σ.**

Partitioning coefficient, $D_U$, combines the relative U concentration in both the metal and silicate phases, and therefore reflects better the migration behavior of U between the metal and silicate phases. However, for three of the run products, their silicate phases were lost during cutting and polishing. Hence a silicate average U concentration at close P, T conditions was used in runs 198, 232, 234 due to the fact that the starting materials are the same in each group. From Fig.7, it is evident that their variation trends with P and T are similar to the trends observed in Fig.6 in each group.

A comparison between the results of this work and those of previous studies are illustrated in Fig.8. The $D_U$ values in this study for Fe-10wt% S (0.003-0.135) and Fe-35wt% S (0.003-0.075) overlap the wide range of $D_U$ values ($D_U$ =0.002-2) determined by Murrell and Burnett (1986) at P ≤ 1.5 GPa. The $D_U$ values for pure Fe in this study (0.003 - 0.09) overlap the range of $D_U$ values (0.001- 0.012) determined by Malavergne et al. (2005). The decreasing $D_U$ with increasing P using the same starting composition in their study may originate from their low experimental T (at 15 GPa, their run T is 1900 $^o$C, or 400 $^o$C lower than our closest P run products). The small $D_U$ values (< 0.001) in the run products of Wheeler et al. (2006) may have originated from their large log fo$_2$ (-2) as indicated in Fig.8, since log fo$_2$ value is considered to have a negative correlation with $D_U$ value in the metal phases (Furst et al., 1982; Murrell and Burnett, 1986; Malavergne et al., 2005).

The results presented here show that the U concentrations and $D_U$ values in all three groups of metal phases increase with increasing experimental P, and they increase at a faster rate when the experimental T is above silicate melting than if the experimental T is below silicate melting. These results suggest that with increasing experimental P, U gradually migrates from silicate or uraninite included in silicate, to the metal phases. When silicate is melted, more U migrates from the silicate to metal phases.

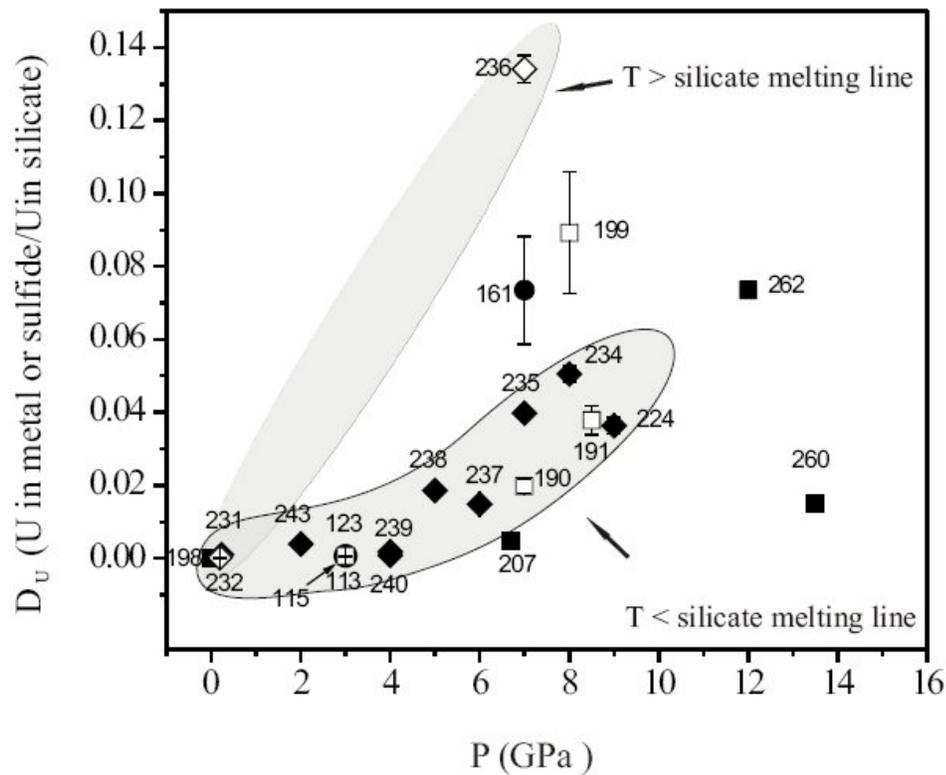

**Fig.7 Partitioning coefficients of U ($D_U$) as a function of pressure. The lower shading region outlines the Fe-10wt% S group with a T below silicate melting. Legends are same as Fig.6. Errors are 2σ.**

The results presented in our previous study (Bao et al., 2006) showed that Si in the pure Fe phase increases with increasing P. Using the relationship between Si concentration in the metal phase and oxygen fugacity (Kilburn and Wood, 1997), it was concluded that increasing P causes a decrease in oxygen fugacity ($f_{O2}$) in the experiments. This can make U deviate from its normal behavior to become less lithophile and enter the metal phase (Murrell and Burnett, 1986). Based on the results presented in Figs. 6 and 7, where U and $D_U$ increase with P, this conclusion also applies for the Fe-10wt% S and Fe-35wt% S run products. Coupled with the effects of P on $f_{O2}$ is the effect of U mobility in molten silicate. The release of U from the silicate fraction due to the high T and the lowering of $f_{O2}$ enhances U affinity for the metal phase. This could be described by the following reactions:

$$UO_2 + 2Fe \leftrightarrow 2FeO + U \quad or: \quad UO_2 + 2FeS \leftrightarrow US_2 + 2FeO$$

In addition, increasing T also causes reduction (Righter and Drake, 2006). This also may explain when T was over silicate melting, more U entered the metal phase based on the relation between U solubility in the metal phase and $f_{O_2}$ discussed above.

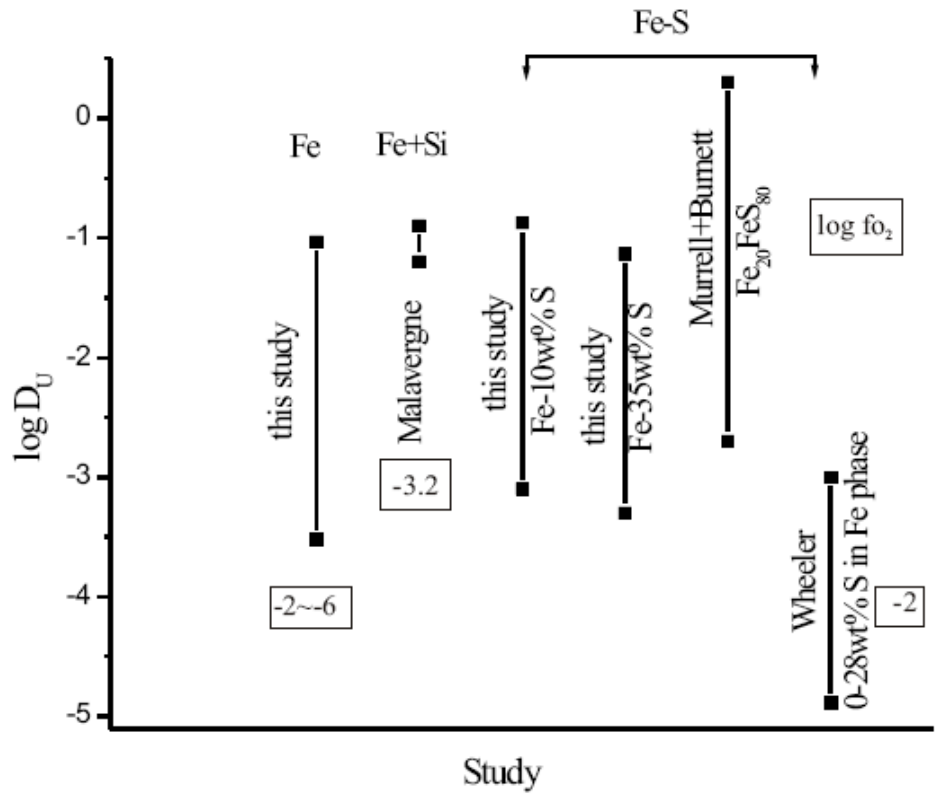

**Fig.8 Comparison between the $D_U$ values achieved by different research groups. The composition of the metal phases in the starting materials and the oxygen fugacity, log fo₂ value of these experiments have been marked in this figure.**

However, in the Fe-10wt% S and Fe-35wt% S run products, the positive correlation between Si concentration and P, as it occurs in pure Fe, was not observed in this study. In fact, within the general scatter in the data, it is found that the Si concentration decreases with increasing P in most of the Fe-10wt% S run products based on the data from the LA-ICP-MS analysis as shown in Fig.9. Si and S are mutually incompatible in the metal phase (Poirier, 1994; Kilburn and Wood, 1997). Therefore, the incorporated S in the metal-sulfide phase may destroy the positive correlation between Si concentration and P observed in the pure Fe phase.

The U versus Ca relation in metal phases is illustrated in Fig.10. Although there is no strongly defined trend, most data tend to show that Ca and U have a weak negative correlation. Herndon (1996) hypothesized that U can be incorporated in the core with CaS. McDonough (2003) argued that if this happened, then the Ca/Ti and Ca/Al ratios of the mantle should be significantly lower than those of chondritic meteorites. However, in fact this is not observed (McDonough and Sun, 1995). The lack of correlation or slightly negative correlation between U and Ca suggested in Fig.10 implies that U can enter the metal phase and the core without Ca accompanying it. This would explain even if U entered the core, the mantle Ca/Ti and Ca/Al ratios remain similar to those of the chondritic meteorites.

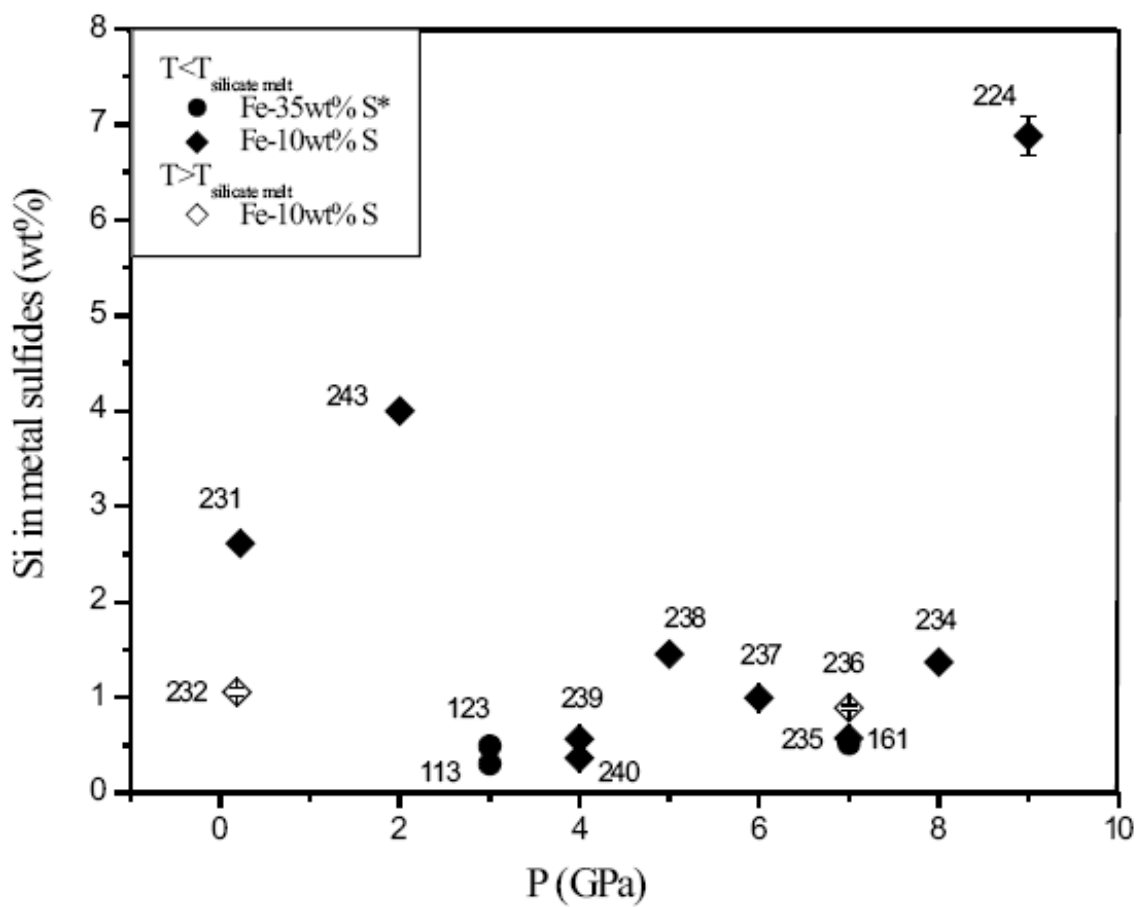

**Fig.9 Si in metal sulfide phases as a function of pressure. Errors are 2σ.    * Containers in this group are graphite, in the others are BN.**

The observation that S enhances U solubility in the metal phase suggests that U may possibly combine with S and exist as US in metal phase. Similarly, Ca may also exist as CaS in metal sulfides. The slightly negative correlation between U and Ca implies that when U increases, it may substitute for Ca in the CaS, and Ca will gradually leave the metal phase. With increasing P, U gradually increases in the metal phases as illustrated in Figs.6 and 7, consequently it is expected that Ca will gradually decrease.

This can be understood from their atomic radius ratios with Fe: $r_U/r_{Fe}$ =1.22, $r_{Ca}/r_{Fe}$=1.48. Compared with Ca, the atomic radius of U is closer to that of Fe, the main atom of the metal phase and the core. Therefore, when P increases, Fe-Fe distance will decrease and the metal phase will prefer to incorporate smaller U rather than Ca. However, their atomic radius ratios with Fe may change under higher P due to exact nature of the compressibilities of Fe, U and Ca. Therefore, the direct applicability of this conclusion to the core needs to be confirmed by higher P experiments.

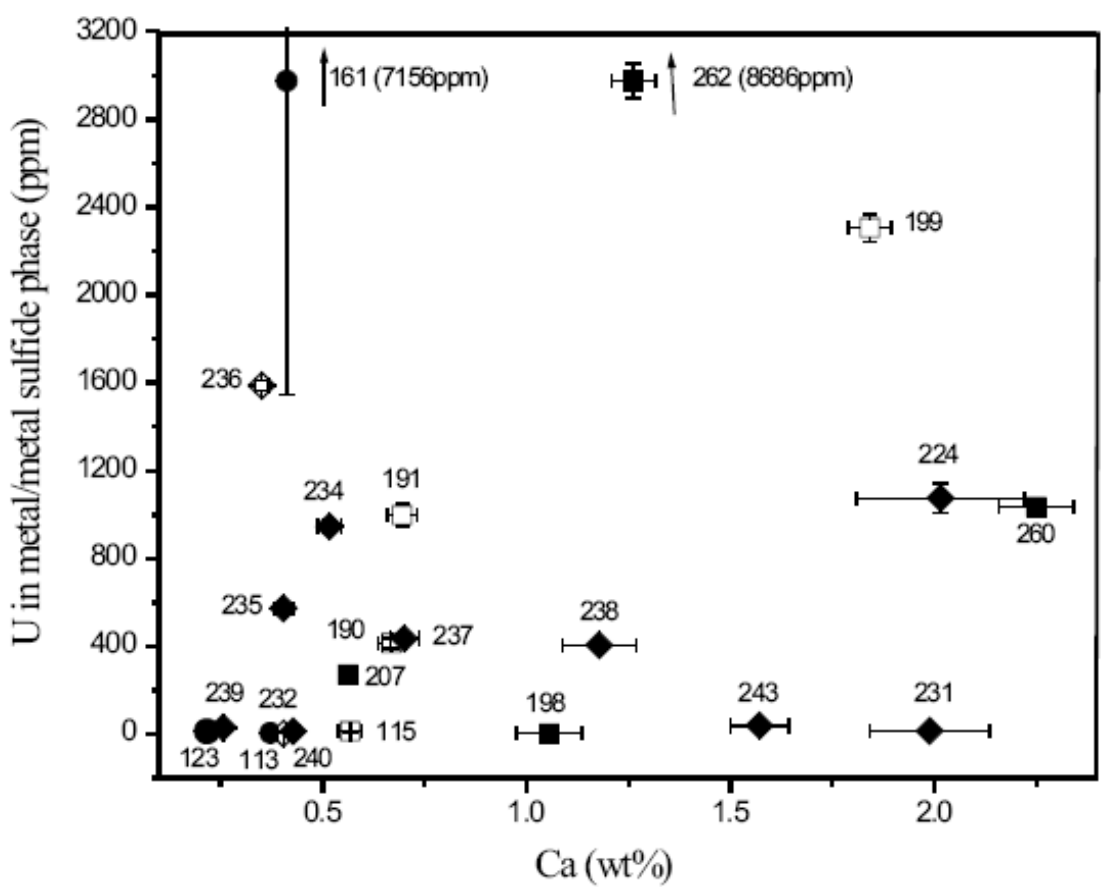

**Fig.10 Ca vs. U concentrations in the Fe and metal-sulfide phases. Legends are same as Fig.6. Errors are 2σ.**

4. IMPLICATION FOR RADIOACTIVE HEATING IN EARTH'S CORE

The Earth's core is currently thought to have formed either from a magma ocean at an early stage (e.g. Righter, 2003) or by percolation (e.g. Bruhn et al., 2000). The percolation model requires a molten metal phase in a solid silicate matrix during core formation. The experiments in this study with temperatures below the silicate melting (see Figs. 6 and 7) are aligned with this model. In the Earth's mantle, P and T increase with depth. At depths where the T exceeds the metal phase melting, liquid metal drops surrounded by silicate matrix will form an interconnected phase and then percolate downward to form the core. The metallic Fe in the whole mantle or lower mantle (Shannon and Agee, 1998) can enter the core by percolation. For comparison with the magma ocean model that will be discussed in the following, the extrapolated P value will be set at 26 GPa, as shown in Fig.11, which is equivalent to the P at the base of an intermediate magma-ocean (Righter, 2003).

From Figs. 6 and 7, it is observed that both the U concentration and $D_U$ in the T < silicate melting line sub-group increases with P, with a steeper increase above 4 GPa. A $D_U$ value of 0.25 is obtained when extrapolated to 26 GPa for Fe-10wt% S group as shown in Fig.11. If the silicate Earth model of 20 ppb U (McDonough, 2003) is used as the average U concentration of the silicate matrix contacting with metal drops in the mantle, then 5 ppb U in the core is obtained using a $D_U$ value of 0.25. This is 5 times larger than the 1ppb value that the pure Fe results for temperature below silicate melting extrapolated to 26 GPa. This further indicates that if S has entered the core as a light element, this should have enhanced the U concentration in the core.

The magma ocean model requires that both silicate and metal phase were melted during core formation, which is similar to the experiments in this study with temperature in excess of silicate melting. There are two sub-models (e.g. Righter, 2003): one is the intermediate depth magma ocean with a P of ~ 26 GPa at its bottom, and the another is deep magma ocean with a P of ~ 50 GPa at its bottom.

Using parallel patterns of behavior between pure Fe and Fe-10wt% S run products for T below silicate melting implies that similar behavior for T above silicate melting can be expected. Therefore, the greater rate of $D_U$ variation in the Fe-10wt% S data at T above silicate melting may be reliable, although there are only two run products in this sub-group. When $D_U$ is extrapolated to a P at the bottom of an intermediate depth magma ocean of 26GPa, a $D_U$ value of 0.51 is obtained as shown Fig.11. This corresponds to the core containing ~10 ppb U, which is 4.3 times larger than the results from the pure Fe study.

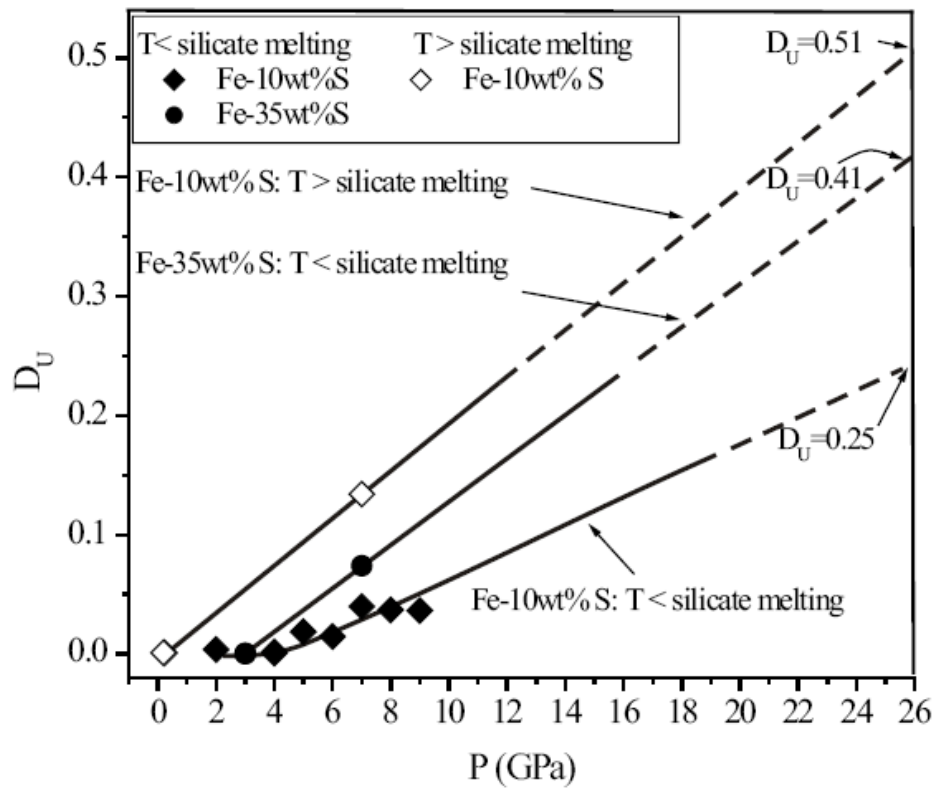

**Fig.11 The $D_U$ values of metal sulfides from different temperature groups were extrapolated to 26 GPa, the pressure at the bottom of a proposed magma ocean.**

Comparing the extrapolated results from both sets of temperatures, $D_U$ values in Fe-10wt% S run products in both temperature levels are found to be 4.3 to 5.0 times larger than the corresponding $D_U$ values of pure Fe run products for the corresponding T sub-groups. It is noted that those comparative values are very close to one another suggesting consistency. This shows that the influence of S in enhancing U solubility in the metal phase in both temperature regions is similar.

In order to illustrate better the influence of S on the U solubility in the metal phase, the $D_U$ values from all three groups extrapolated to 26 GPa are plotted against S content as shown in Fig.12. It clearly indicates that the $D_U$ value increases with increasing S content in the metal components of the starting materials for both T levels.

From the timing and rate of inner core solidification, Labrosse et al. (2001) argued that if there are no radioactive elements in the core, then the inner core age is < 1.5 Ga, based on its present radius. However, based on [186]Os-[187]Os geochemical constraints, Brandon et al. (2003) predicted that the inner core should have begun to crystallize prior to 3.5 Ga, which is consistent with the geomagnetic rock record, if the operation of the geomagnetic field depends on the existence of an inner core (e.g. Brodholt and Nimmo, 2002). Therefore, there is a need for some radioactive elements in the core to slow down the rate of temperature decrease and concomitant crystallization of the inner core with time (Labrosse et al., 2001). They further agued that if U is the only heat-producing element in the core, in order to have an inner core age greater than 2.7 Ga, the concentration of U in the core should be ≥ 10 ppb. From the extrapolated result of 10 ppb U in the core from Fe-10wt% S run products, an inner core age greater than 2.7 Ga can be explained, if the core formed from an intermediate magma ocean and the core contains 10 wt % S. Alternatively, if the core formed by percolation, the extrapolated result at 26 GPa is 5 ppb U in the core. Because U in the deep mantle can still percolate to the core, therefore the core should contain more than 5 ppb U (possibly 22 ppb when $D_U$ is extrapolated to Core Mantle Boundary (CMB)). Hence, if the core contains 10 wt% S, the results of this study suggest that regardless of how the core formed, from a magma ocean or by percolation, the U concentration in the core may contribute an explaination, in whole or in part, for an inner core age greater than 2.7 Ga.

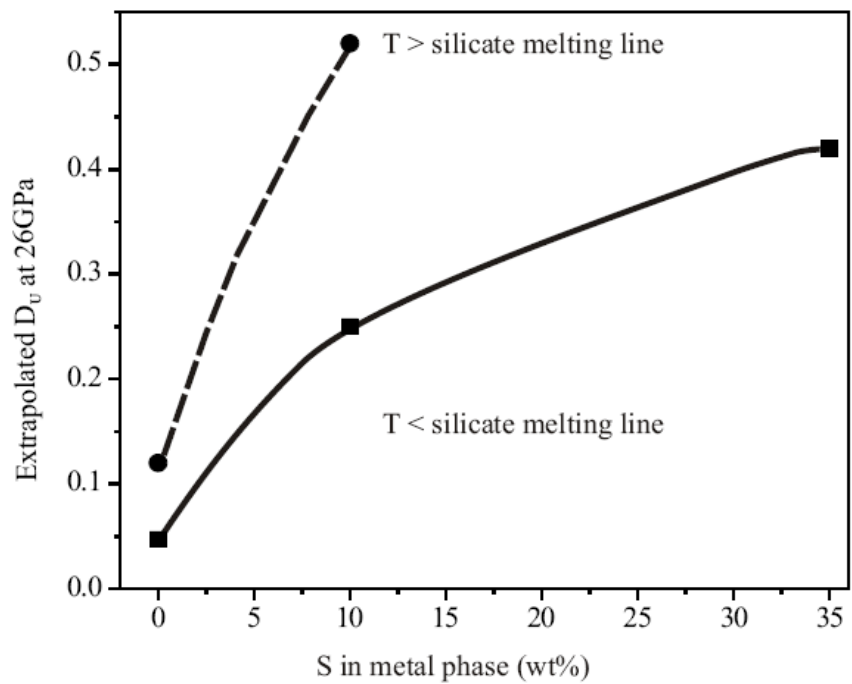

**Fig.12 Influence of S concentration in metal phases on the extrapolated $D_U$ values at 26 GPa, the pressure at the bottom of a proposed magma ocean.**

On the other hand, the existence of an inner core may not be a prerequisite for the generation and maintenance of a dipolar geomagnetic field (Butler et al., 2005). In this case, radioactive elements (Butler et al., 2005), such as U existing in the core prior to 3.5 Ga will be a pre-condition for offering energy to drive the Earth's heat engine and producing a geomagnetic field.

For a detailed discussion of the implications for terrestrial planetary dynamics, please refer to Bao (2006).

## 5. CONCLUSIONS

Based on the experimental results presented, the following conclusions are made:

From pure Fe, Fe-10wt% S to Fe-35wt% S run products, U solubility in the metal phases increases with increasing S concentration as summarized in Fig.12.

The results on Fe-S confirm the conclusion based on the pure Fe system that U solubility in metal phases increases with P and T. As in those studies, it is also found that when the experimental T is above silicate melting, $D_U$ is 3-6 times larger that that of the low T run products.

If Earth's core formed from an intermediate magma ocean and contains 10wt% S, then it could contain approximately 10 ppb U and could have an inner core greater than 2.7 Ga. If it formed from a deep magma ocean and contains 10wt% S, then it is expected to have more than 10 ppb U in the core.

Alternatively, if the core formed by percolation, then it is expected that the core can contain more than 5 ppb U based on the limited (26GPa) extrapolation of $D_U$. However, the percolation model predicts that the Fe in the whole lower mantle can percolate to the core, therefore, the $D_U$ should be extrapolated to the CMB pressure (135 GPa) –highlighting the need for experimental data at higher P. Therefore, it is expected that the core can contain much higher U than 5 ppb (probably 22 ppb) and possibly satisfy an inner core greater than 2.7 Ga.

**Acknowledgments**

This work was supported by a grant awarded by the National Sciences and Engineering Research Council of Canada to RAS. We thank G. Young for his partial funding support to XB. C. Cermignani, M. Liu and L. Shi for electron microprobe analysis; W. Minarik for LA-ICP-MS analysis. A. Pratt for his help with U analysis; G. Gord for SIMS analysis; D. Liu for his help in sample preparation; R. Tucker for his help in preparing high pressure components and G. Wood for his help in sample cutting and polishing.